\documentclass[a4paper, 12pt]{article}
\usepackage[utf8]{inputenc}
\usepackage[T1]{fontenc}
\usepackage{amsmath,amssymb, amsfonts}
\usepackage{hyperref}
\usepackage{icomma}
\usepackage{color}
\usepackage{graphicx}
\usepackage{hyperref}
\usepackage{titlesec}
\usepackage{lmodern}
\usepackage{units}
\usepackage{marvosym}
\usepackage[usenames,dvipsnames,svgnames,table]{xcolor}
\usepackage{cite}
\usepackage{xypic}
\usepackage{marginnote}
\usepackage[marginparwidth=2.0cm]{geometry}
\usepackage{subcaption}

\begin{document}
\title{ Off-shell structure of twisted (2,0) theory}
 \author{Ulf Gran, Hampus Linander, Bengt E.W. Nilsson}
 \date{}
\maketitle
\begin{center}

\vspace*{-1em}
Department of Fundamental Physics\\
Chalmers University of Technology\\
S-412 96 G\"oteborg, Sweden\\[3mm]
{\tt ulf.gran@chalmers.se, linander@chalmers.se, tfebn@chalmers.se}
\end{center}

\vspace{1em}
\begin{abstract}
  A $Q$-exact off-shell action is constructed for twisted abelian (2,0) theory on a Lorentzian six-manifold of the form $M_{1,5} = C\times M_4$, where $C$ is a flat two-manifold and $M_4$ is a general Euclidean four-manifold. The properties of this formulation, which is obtained by introducing two auxiliary fields, can be summarised by a commutative diagram where the Lagrangian and its stress-tensor arise from the $Q$-variation of two fermionic quantities $V$ and $\lambda^{\mu\nu}$. This completes and extends the analysis in \cite{twisted-2-0}.
\end{abstract}

\newpage

\tableofcontents

\section{Introduction}
In this note we consider twisted (2,0) theory on a Lorentzian six-manifold  of the form $M_{1,5}=C\times M_4$ where $C$ is a flat, Lorentzian two-manifold \cite{twisted-2-0}\footnote{The twisting is carried out in Minkowski signature where the non-compact part of the Lorentz group prevents a full twist but where a low energy limit still produces supercharges with the required properties for a topological theory.}. This setup is interesting since it could give some insight into the conjectured correspondence between four-dimensional gauge theory and two-dimensional CFT known as the AGT-correspondence \cite{Yagi:1, AGT}. This is one part of a larger web \cite{Gaiotto:N2dualities, Witten:2011, Dimofte:2011ju, Kapustin:2006pk, vafa-witten} of dualities and relations that can be derived assuming the existence of the elusive superconformal theory in six dimensions known as (2,0) theory\cite{Witten:1995, Witten:ct46}.

In previous work \cite{twisted-2-0} the twisted theory on $M_4$ was calculated explicitly in terms of the free tensor multiplet. It was shown that on a flat background there is a $Q$-exact and conserved stress tensor but that these properties did not immediately  extend to a general curved $M_4$. The  problem  was located to the stress tensor for the bosonic self-dual two-form which turned out not to be conserved on a general four-manifold.
However, this issue needs to be remedied since the procedure of topological twisting should  result in theories with $Q$-exact stress tensors defined on a generally curved background \cite{witten:1988, yamron:1988, marcus:1995}.

Here we construct an action for the full theory that is $Q$-exact off-shell using two different kinds of auxiliary field. The free theory splits into two parts, one of which is equivalent to Donaldsson-Witten theory, and hence this sector can be taken off-shell following the standard techniques described in \cite{Baulieu:1988, Brooks:1988, Birmingham:1991}. 

In the other sector there is a self-dual tensor field whose presence in the $Q$ transformation rules leads to an unwanted metric dependence. However, by the introduction  of an auxiliary vector field we are able to eliminate this metric dependence  in a similar fashion as in \cite{lee-lee-park}. Also in this sector, this step leads to a formulation where the scalar supercharge is nilpotent off-shell and after constructing a $Q$-closed and  covariant Lagrangian for the entire theory
these properties become manifest also for the stress-tensor. For the bosonic self-dual two-form we have also  added certain curvature terms to its equation of motion,
which becomes possible  after the twisting is  performed. Note that  these terms cannot  be obtained in the original (2,0) theory in six dimensions but they are known from  the closely related interacting theory  constructed  in five dimensions in \cite{ Witten:2011}.

With these modifications of the theory that is naively obtained from twisting the six-dimensional (2,0) theory on $M_4$, we find an off-shell theory whose metric variations and $Q$ transformations commute. This feature then implies that the stress tensor can be derived from a fermionic quantity $V$ (given below) either by going via the Lagrangian or via $\lambda^{\mu\nu}$ (where $T^{\mu\nu}=\{Q,\lambda^{\mu\nu}\}$). In section 4 this is summarized in a {\it commuting square} whose corners  represent the four  quantities involved, i.e., $T^{\mu\nu}$, $\lambda^{\mu\nu}$, $V$ and its $Q$ transform, the Lagrangian.

In section \ref{sec:twist} we review the four-dimensional theory obtained by twisting the six-dimensional (2,0) theory on $C\times M_4$. The problem encountered previously and its resolution are briefly explained in section \ref{sec:conservation}. In section \ref{sec:Q-exact} we construct an off-shell formulation including a $Q$-exact action. Finally, in section \ref{sec:conclusions} we summarise and comment on the results.

\section{The twisted theory}
\label{sec:twist}
For the convenience of the reader we here give a short review of the twisted theory, for details see \cite{twisted-2-0}. On a general background the six-dimensional $(2,0)$ theory admits no twist that preserves any supersymmetry since the $\mathrm{Spin}(5)$ $\mathrm{R}$-symmetry cannot be used to fully twist the supercharges transforming in the larger six dimensional Lorentz group $\mathrm{Spin}(1,5)$. However, on specific backgrounds such as the one considered here of the form $M_6=C\times M_4$, the Lorentz group is small enough. Here we twist by considering a new $\mathrm{SU}(2)'$ as the diagonal embedding
\begin{equation}
  \mathrm{SU}(2)'=\mathrm{SU}(2)_{\mathrm{r}}\times \mathrm{SU}(2)_\mathrm{R}\,,
\end{equation}
where the six dimensional Lorentz group is $\mathrm{Spin}(1,1)\times \mathrm{SU}(2)_\mathrm{l}\times\mathrm{SU}(2)_\mathrm{r}$ and the $\mathrm{R}$-symmetry subgroup is given by 
\begin{equation}
  \mathrm{SU(2)}_\mathrm{R}\times \mathrm{U}(1)_\mathrm{R} \cong \mathrm{Spin}(3)\times \mathrm{Spin}(2) \subset \mathrm{Spin}(5)_\mathrm{R}.
\end{equation}

The supercharges transform in the $(\pmb{1}, \pmb{2}, \pmb{2}^{\pm})$ of $\mathrm{SU}(2)_\mathrm{l}\times \mathrm{SU}(2)_\mathrm{r} \times \mathrm{SU}(2)_\mathrm{R}\times \mathrm{U}(1)_\mathrm{R}$ which after twisting results in two scalar supercharges on $M_4$ of which we pick the one with negative $\mathrm{U}(1)_\mathrm{R}$ charge\footnote{This charge corresponds to the one that would become scalar on $C$ under the full twisting in the Euclidean scenario\cite{twisted-2-0}.}. This charge satisfies $Q^2=0$ and if one also finds a $Q$-exact stress tensor the theory is topological on $M_4$\cite{witten:1988}.

The scalars $\Phi$ of (2,0) theory transform in the vector $\pmb{5}$ of $\mathrm{Spin}(5)_\mathrm{R}$ and thus after twisting consist of one self-dual two-form $E_{\mu \nu}$ and one complex scalar $\sigma$. The symplectic Majorana-Weyl spinor $\Psi$ after twisting contains two sets of fields with opposite $U(1)_\mathrm{R}$-charge, all of which are Grassmann. The first set consists of a scalar $\eta$, a one-form $\psi_\mu$ and a self-dual two-form $\chi_{\mu\nu}$. The second set is a copy of the aforementioned one with opposite $\mathrm{U}(1)_\mathrm{R}$-charge, denoted with a tilde. The self-dual three-form $H$ gives rise to a one-form $A$ and a two-form $F$ which we split into its self-dual and anti-self-dual parts $F^+$ and $F^-$. 

The equations of motion of the twisted theory after reduction to a flat $M_4$ is given by

\begin{equation}
  \begin{aligned}
    \partial_\rho \partial^\rho E_{\mu \nu} &= 0 \\
    \partial_\rho \partial^\rho \sigma &= 0 \\
    \partial_\rho \partial^\rho \bar{\sigma} &= 0 \\
  \end{aligned}
  \qquad
  \begin{aligned}
    \partial_{[\mu} A_{\nu]} &= 0 \\ 
    \partial_{[\mu} F^\pm_{\nu \rho]} &= 0 \\ 
    \partial_\mu A^\mu &= 0
  \end{aligned}
  \qquad
  \begin{aligned}
    \partial_\mu \tilde{\psi}^\mu &= 0 \\
    \partial_\mu  \tilde{\eta} - \partial_\nu \tilde{\chi}_{\mu}{}^{\nu} &= 0 \\
    (\partial_{[\mu}  \tilde{\psi}_{\nu]})^+ &= 0 
  \end{aligned}
  \qquad
  \begin{aligned}
    \partial_\mu \psi^\mu &= 0 \\
    \partial_\mu  \eta - \partial_\nu \chi_{\mu}{}^{\nu} &= 0 \\
    (\partial_{[\mu}  \psi_{\nu]})^+ &= 0 
  \end{aligned}
\end{equation}
where the notation $(\dots)^+$ refers to the self-dual part.

This set of equations is invariant under the supersymmetry transformations
\begin{equation}
  \begin{aligned}
    \delta E_{\mu  \nu} & =  i  \chi_{\mu  \nu} v \\
    \delta \tilde{\psi}_\nu &= i v  A_\nu-  v   \partial_\mu E_{\nu}{}^{\mu} \\
    \delta  A_{\mu} & =   \partial_{\mu} \eta \\
    \delta \chi_{\mu \nu} &= 0\\ 
    \delta \eta &= 0
  \end{aligned}
  \qquad
  \begin{aligned}
    \delta F_{\mu \nu}^+ & =  0 \\
    \delta  F_{\mu \nu}^- & =   - 4 \partial_{\left[ \mu \right.} \psi_{\left. \nu \right]} v\\
    \delta \tilde{\chi}_{\mu \nu} &= 2 i v F_{ \mu \nu}^+ \\
    \delta \psi_\nu &= - v  i \sqrt{2}\partial_\nu \bar{\sigma} \\
    \delta \sigma & = \sqrt{2} \tilde{\eta} v \\
    \delta \bar{\sigma} & =  0 \\
    \delta \tilde{\eta} &= 0
  \end{aligned}
\label{eq:twistedsusy}
\end{equation}
where $v$ is a Grassmann parameter. As written the transformations in the left hand column close on-shell using the equation of motion $\partial_\mu \eta - \partial_\nu \chi_\mu{}^\nu$.

The free theory splits into two sectors corresponding to the two columns in \eqref{eq:twistedsusy}. The first consists of $\{ E_{\mu\nu}, \tilde{\psi}_\mu, A_\mu, \chi_{\mu\nu}, \eta\}$, henceforth called the $E$-sector, and the second containing the Yang-Mills field strength $\{ F_{\mu\nu}, \tilde{\chi}_{\mu\nu}, \psi_\mu, \sigma, \bar{\sigma}, \tilde{\eta} \}$, referred to as the $F$-sector. The latter sector correspond to the field content of Donaldsson-Witten theory, i.e.\ the unique twist of pure ${\cal N}=2$ supersymmetric Yang-Mills. Note that the supersymmetry transformations also corresponds to Donaldsson-Witten theory except for $\delta F$, a point to which we will return  in section \ref{sec:Q-exact}. The former sector does not stem from any ${\cal N}=2$ multiplet\footnote{An ${\cal N}=2$ hypermultiplet would result in one bosonic and two fermionic vectors.}. However, it is closely related to the Vafa-Witten twist\cite{vafa-witten} of ${\cal N}=4$ supersymmetric Yang-Mills and the topological twisting of five-dimensional supersymmetric Yang-Mills\cite{Witten:2011, Anderson:2012}. These connections stem from the fact that both the four- and five-dimensional untwisted theories can be obtained by compactifications of (2,0) theory\cite{Seiberg:noteon16}.

\section{Conserved stress-tensor}
\label{sec:conservation}

In \cite{twisted-2-0} it was shown that there exists a conserved and $Q$-exact stress tensor on a flat background, which we here split into the two sectors defined in the previous section. 
It is then given by the sum of the two stress tensors 
\begin{equation}
  T^{\mu\nu}_i = \{Q, \lambda^{\mu\nu}_i\}\,,
\end{equation}
where
\begin{equation}
  \label{eq:lambda1}
  \begin{aligned}
    \lambda^{\mu \nu}_1 &=
    \frac{1}{2}\Big(
     \tilde{\psi}^{(\mu} \partial^\rho E^{\nu)}{}_{\rho} 
    + \partial_{\rho}\tilde{\psi}^{(\mu} E^{\nu)}{}_{\rho} 
    - \partial^{(\mu}\tilde{\psi}^{\rho}  E^{\nu)}{}_{\rho}
    \\ &\phantom{=} 
    + i \tilde{\psi}^{(\mu} A^{\nu)}
    -\frac{1}{2}g^{\mu \nu}\tilde{\psi_\rho} \partial_\sigma E^{\rho \sigma}
    - \frac{i}{2} g^{\mu \nu} \tilde{\psi}_\rho A^\rho \Big)\,,
  \end{aligned}
\end{equation}
and
\begin{equation}
  \begin{aligned}
    \label{eq:lambda2}
    \lambda^{\mu \nu}_2 &=
    \frac{1}{2}\Big(
    \sqrt{2}i \psi^{(\mu} \partial^{\nu)} \sigma
    - \frac{i}{2}  \tilde{\chi}^{(\mu}{}_\rho  F^-{}^{ \nu)}{}^{\rho} 
    -\frac{i}{\sqrt{2} }g^{\mu \nu}\psi_\rho \partial^\rho \sigma \Big)
    .\end{aligned}
\end{equation}

Since any stress tensor on a curved background must reduce to the above in the flat limit a natural guess for the curved stress tensor is by covariantising $\lambda^{\mu\nu}_i$. Furthermore, by the symmetries of the theory there are no curvature corrections that can be added to $\lambda^{\mu\nu}_i$ and since the terms in $\lambda^{\mu\nu}_i$ have only one derivative no new curvature factors can arise by ordering (see \cite{twisted-2-0} for details).

Using a covariantised version of $\lambda^{\mu \nu}$ one finds the stress tensor
\begin{equation}
\begin{aligned}
\label{eq:Stress_Tensor_curvedE}
  T^{\mu \nu}_1 &= \frac{1}{8}\left( -4 A^{(\mu} A^{\nu)} + 
2 g^{\mu \nu} A_\rho A^\rho  \right)
+ 
 \frac{i}{2} g^{\mu \nu}
  D_\rho \eta \tilde{\psi}^\rho 
- i    
 D^{(\mu} \eta \tilde{\psi}^{\nu)} 
\\ & \phantom{=}
- \frac{i}{4} g^{\mu \nu}
    \chi^{\rho \sigma}  D_{[\rho}  \tilde{\psi}_{\sigma]}
+\frac{i}{2} \left(
\chi^{\sigma (\mu}D_\sigma \tilde{\psi}^{\nu)} 
-
\chi^{\sigma (\mu } D^{\nu)} \tilde{\psi}_{\sigma}
\right) 
\\ 
& \phantom{=}
+ \frac{1}{4}g^{\mu \nu}     D^\kappa E_{\rho \kappa}    D_\sigma E^{\rho \sigma}
-  \frac{1}{2}   D^\rho \Big(  D^\kappa E^{(\mu}{}_{\kappa}  E^{\nu)}{}_{\rho} \Big)
+ \frac{1}{2}  D^{(\mu}    D_\kappa E^{\rho \kappa}  E^{\nu)}{}_{\rho}
\end{aligned}
\end{equation}

\begin{equation}
\begin{aligned}
\label{eq:Stress_Tensor_curvedF}
  T^{\mu \nu}_2 &= \frac{1}{2} \Big(
-g^{\mu \nu} D_{\rho} \sigma D^\rho \bar{\sigma} + 2 D^{(\mu} \sigma D^{\nu)} \bar{\sigma} 
\Big)
 +\frac{1}{8}\left( - 2F^+{}^{\mu}{}_{\rho}F^-{}^{\rho\nu} - 2 F^-{}^{\mu}{}_{\rho}F^+{}^{\rho\nu} \right)
\\ 
&\phantom{=}+ 
 \frac{i}{2} g^{\mu \nu}
 D_\rho  \tilde{\eta} \psi^\rho
- i 
 D^{(\mu}  \tilde{\eta} \psi^{\nu)}
- \frac{i}{4} g^{\mu \nu}
  \tilde{\chi}^{\rho \sigma}   D_{[\rho}  \psi_{\sigma]}
  \\ & \phantom{=}
+\frac{i}{2} \left(
\tilde{\chi}^{\sigma (\mu}D_\sigma \psi^{\nu)} 
-\tilde{\chi}^{ \sigma (\mu  } D^{\nu)} \psi_{\sigma}
\right) \,.
\end{aligned}
\end{equation}

It turns out that the part of this stress tensor involving the bosonic self-dual two-form (last line in \eqref{eq:Stress_Tensor_curvedE}) is not conserved using the equations of motion obtained by twisting the six-dimensional equations of motion,

\begin{equation}
  D^2 \Phi = 0 \,,
\end{equation}
which implies that $E_{\mu\nu}$, $\sigma$ and $\bar{\sigma}$ satisfies the corresponding four-dimensional equations. 

One natural guess is that since the six-dimensional theory on a curved background is conformally invariant only when the conformal coupling $R\Phi^2$ is included, it will generate the needed terms in the twisted theory. It turns out that this is not enough\cite{twisted-2-0} since an addition to the equation of motion of the form
\begin{equation}
  D^2 E_{\mu \nu} = a R E_{\mu \nu}
\end{equation}
does not enable the conservation of \eqref{eq:Stress_Tensor_curvedE} for any value of $a$. 

In fact, after twisting, there are other possible curvature couplings allowed by the index structure and symmetry. 
Such terms do not arise by twisting any six-dimensional expressions but as we will show they solve the problem of conservation and also have a very simple Lagrangian description.

The Ricci tensor gives the possibility to add to the equations of motion a term of the form
\begin{equation}
  D^2 E_{\mu\nu} = b (P^+){}_{\mu \nu}{}^{\tau \sigma}R_{\tau}{}^{\rho} E_{\sigma\rho}\,,
\end{equation}
where $P^+$ is the projector on the self-dual part. It turns out that this projection is actually proportional to the curvature scalar and so does not contribute anything new. 
The last possible addition is a term of the form
\begin{equation}
  D^2 E_{\mu \nu} = c (P^+){}_{\mu \nu}{}^{\tau\sigma}R_{\tau \sigma}{}^{\rho \lambda} E_{\rho \lambda}\,.
\end{equation}

It is now a simple matter to confirm that the stress tensor in \eqref{eq:Stress_Tensor_curvedE} is conserved with the additions to the equations of motion given by $a=\frac{1}{2}$, $c=-1$. The correct equation of motion is then given by
\begin{equation}
  D^2 E_{\mu\nu} = \frac{1}{2} R E_{\mu \nu} - (P^+){}_{\mu \nu}{}^{\tau\sigma}R_{\tau \sigma}{}^{\rho \lambda} E_{\rho \lambda} \,.
\end{equation}
This equation integrates to the Lagrangian\footnote{This Lagrangian coincides with the corresponding part of the one derived previously from a five-dimensional perspective\cite{Witten:2011, Anderson:2012} after reduction to four dimensions.}.
\begin{equation}
  L_E = E^{\mu \nu} D^\rho D_\rho E_{\mu \nu} + R^{\mu \nu \rho \sigma} E_{\mu \nu} E_{\rho\sigma} - \frac{1}{2} R E^{\mu \nu} E_{\mu \nu}\,,
\end{equation}
which after a partial integration\footnote{In the absence of boundary terms.} is equivalent to the simpler form
\begin{equation}
  L_E = -4 D^\mu E_{\nu \mu} D_\rho E^{\nu \rho}.
  \label{eq:LE}
\end{equation}
From here one can also easily verify that the stress tensor in \eqref{eq:Stress_Tensor_curvedE} follows from a metric variation of \eqref{eq:LE}, keeping in mind the metric dependence of $E_{\mu \nu}$ due to its self-duality (see the appendix for details).

\section{$Q$-exact action}
\label{sec:Q-exact}
After having found a $Q$-exact stress-tensor a natural question is if the action itself is $Q$-exact. It turns out that this is the case but that there are a few subtleties. In fact, in the end, we will find a commuting diagram shown in Figure \ref{fig:comdia} for the different sectors. Here the vertical direction corresponds to metric variations and should be considered to take place under an integral in the sense that
\begin{equation}
  \delta_g \int_{M_4}\!\!\sqrt{g}\,V = \int_{M_4}\!\!\sqrt{g}\, \delta g_{\mu\nu} \lambda^{\mu\nu}, 
\end{equation}
as well as the more familiar
\begin{equation}
  \delta_g \int_{M_4} \!\!\sqrt{g}\,L = \int_{M_4} \!\!\sqrt{g}\, \delta g_{\mu\nu} T^{\mu \nu}.
\end{equation}
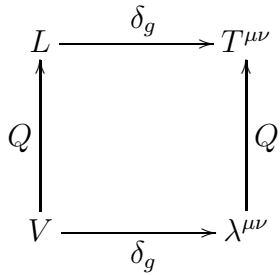
\begin{figure}
  \centering
    \[
      \renewcommand{\labelstyle}{\textstyle}
      \xymatrix@R=2cm@C=2cm{ 
        L  \ar[r]^{\delta_g}  & T^{\mu\nu}\\
         V \ar[r]_{\delta_g} \ar[u]^{Q} &   \lambda^{\mu\nu} \ar[u]_{Q} 
      }
    \]
  \caption{Relation between the fermionic quantity $V$ and the stress tensor. The vertical direction correspond to supersymmetry transformations and the horizontal direction to metric variations.}\label{fig:comdia} 
\end{figure}
First, let us see why we should expect to find the relationship in Figure \ref{fig:comdia} in topological theories of cohomological type.
If a theory has a $Q$-exact action, ${L=\{Q, V\}}$, it follows that 
\begin{equation}
  \begin{aligned}
  \delta_g \int_{M_4} \!\!\sqrt{g}\,L &= \int \!\!\sqrt{g}\left( \frac{1}{2}\mathrm{Tr}(\delta g) \{Q, V\} + \delta_g \{Q, V\}\right) \\
  &= \int \!\!\sqrt{g}\, \delta g_{\mu\nu} \left\{Q,\, \frac{1}{2}\mathrm{Tr}(\delta g)V + \frac{\delta V}{\delta g_{\mu\nu}}\right\},
\end{aligned}
\label{eq:comdiacalc}
\end{equation}
i.e.\ the stress-tensor is $Q$-exact. However, care is needed when performing the above calculation. Note that it breaks down if the supersymmetry variations do not commute with the metric variations. Below we show that this indeed occurs for the $E$-sector. It is also the case that the Lagrangian for the $F$-sector is only $Q$-exact on-shell and so the situation is not as straight forward as in \eqref{eq:comdiacalc}.
\subsection{ $E$-sector }
It turns out that in the twisted theory at hand there is in the variation of $\tilde{\psi}$
\begin{equation}
  \delta \tilde{\psi}_\mu = i v A_\mu - v D_\rho E_{\mu}{}^{\rho}\,,
\end{equation}
a metric dependence in the second term on the right hand side, due both to the covariance and to the self-duality of $E_{\mu\nu}$. Since this term does not have the same  dependence 
on the metric as the other terms it follows that metric variations and supersymmetry transformations do not commute on $\tilde{\psi}_\mu$. 

One of the consequences of this can be seen when trying to construct a fermionic quantity that gives rise to $\lambda^{\mu\nu}$ under a metric variation. 
In the expression for $\lambda^{\mu\nu}_1$ \eqref{eq:lambda1} there is a term of the form $g^{\mu\nu} \tilde{\psi}_\rho D_\sigma E^{\rho \sigma}$. To generate this we look at the corresponding term
\begin{equation}
  V = \tilde{\psi}_\rho D_\sigma E^{\rho \sigma} 
\end{equation}
However, under a metric variation, taking into account the metric dependence of $E^{\mu \nu}$, one finds that
\begin{equation}
  \delta_g\int \!\!\sqrt{g}\, V = \frac{1}{4}\int \sqrt{g}\left( \delta g^{\mu \nu} g_{\mu \nu} D_\sigma \tilde{\psi}_\rho E^{\sigma \rho} + \delta g^{\mu \nu} D_{[\rho} \tilde{\psi}_{\nu]} E_\mu{}^\rho\right).
\end{equation}
Note that no term of the form $\delta g^{\mu\nu} g_{\mu\nu} \tilde{\psi}_\rho D_\sigma E^{\rho \sigma}$ is generated.

The situation can be improved with the aid of an auxiliary field. If we insist on a \textit{uniform metric dependence} in the variation for $\tilde{\psi}$ a natural modification to do is to replace $E$ in this variation by a metric independent auxiliary field $h_\mu$. Thus we now have
\begin{equation}
  \delta \tilde{\psi}_\mu = i v A_\mu + h_\mu.
\end{equation}
To leave the on-shell theory unchanged we take the action for $h_\mu$ to be
\begin{equation}
  L_h = \frac{1}{2} h^2 - h_\mu D_\nu E^{\mu \nu},
\end{equation}
and use the supersymmetry variation
\begin{equation}
  \delta h_\mu = -i v D_\mu \eta.
\end{equation}
With this modification it is straight forward to find
\begin{equation}
  V_1 = -\left( \frac{1}{2} ( i A_\mu + h_\mu) - D_\nu E_\mu{}^\nu \right) \tilde{\psi}^\mu\,,
\end{equation}
that under a supersymmetry variation gives rise to the action
\begin{equation}
  L_1 = \frac{1}{2}A_\mu A^\mu - i \chi^{\mu \nu} D_\mu \tilde{\psi}_\nu + i D_\mu \eta \tilde{\psi}_\mu + \frac{1}{2} h_\mu h^\mu - h_\mu D_\nu E^{\mu \nu}.
\end{equation}
Note that if we evaluate the auxiliary field on-shell the action for $E_{\mu\nu}$ is exactly of the form \eqref{eq:LE}.
Under a metric variation we find
\begin{equation}
\begin{aligned}
  \lambda_1^{\mu\nu} &= 
  \frac{i}{2} A^{(\mu}\tilde{\psi}^{\nu)}-\frac{i}{4} g^{\mu\nu} A^{\rho}\tilde{\psi}_{\rho} - \frac{1}{2} E^{(\mu}{}_\rho D^{\nu)}\tilde{\psi}^{\rho} - \frac{1}{2} E^{\rho (\mu} D_{\rho} \tilde{\psi}^{\nu)} \\
  &\phantom{=} - \frac{1}{4} g^{\mu\nu} E^{\rho \sigma} D_\rho \tilde{\psi}_\sigma + \frac{1}{2} \tilde{\psi}^{(\mu} h^{\nu)} - \frac{1}{4} g^{\mu\nu} \tilde{\psi}^\rho h_\rho.
\end{aligned}
\end{equation}
One easily verifies that on-shell this expression reduces to the corresponding terms in \eqref{eq:lambda1}. 
Thus we now find that the metric variation of $L_1$ and the supersymmetry variation of $\lambda_1^{\mu\nu}$ both give
\begin{equation}
  \begin{aligned}
    T_1^{\mu\nu} &= -\frac{1}{2} A^{\mu}A^{\nu} + \frac{1}{4}g^{\mu\nu} A^\rho A_\rho + \frac{i}{2} \chi^{(\mu}{}_\rho D^{\nu)}\tilde{\psi}^\rho + \frac{i}{2} \chi^{\rho (\mu} D_\rho \tilde{\psi}^{\nu)} - \frac{i}{4} g^{\mu\nu} \chi^{\rho \sigma} D_\rho \tilde{\psi}_\sigma\\
    &\phantom{=} - i D^{(\mu} \eta \tilde{\psi}^{\nu)} + \frac{i}{2} g^{\mu\nu} D_{\rho} \eta \tilde{\psi}^\rho - \frac{1}{2} h^{\mu} h^{\nu} + \frac{1}{4} g^{\mu\nu} h_\rho h^\rho \\
    &\phantom{=} - \frac{1}{2} E^{\rho (\mu} D^{\nu)} h_\rho + \frac{1}{2} E^{\rho (\mu} D_\rho h^{\nu)} - \frac{1}{4} g^{\mu\nu} E^{\rho \sigma} D_\rho h_\sigma 
  \end{aligned}
\end{equation}
which on-shell coincides with \eqref{eq:Stress_Tensor_curvedE}. This use of an auxiliary field to eliminate the metric dependence of $Q$ and enforce its nilpotency off-shell is very similar in spirit to what is carried out in \cite{lee-lee-park}.

\subsection{$F$-sector}
Let us now turn to the sector containing $F^+_{\mu\nu}$ and $F^-_{\mu\nu}$. Recall that they correspond to the components $H_{+\mu\nu}$ and $H_{-\mu\nu}$ of the six-dimensional self-dual three-form. The contribution to the stress tensor of these fields where derived in \cite{twisted-2-0} by considering first the stress-tensor for a general three-form and only later imposing self-duality, fixing the numerical factors by supersymmetry. This approach is forced on us since the self-dual three-form does not have a covariant Lagrangian. As will soon become evident there are also here some subtleties regarding the construction of $V_2$ generating the wanted terms and also making the diagram in Figure \ref{fig:comdia} commute. From the similarities of this sector with Donaldsson-Witten theory \cite{witten:1988} one would expect the action to be $Q$-exact in the same way\cite{spence}. This is indeed the case apart from one subtle point.

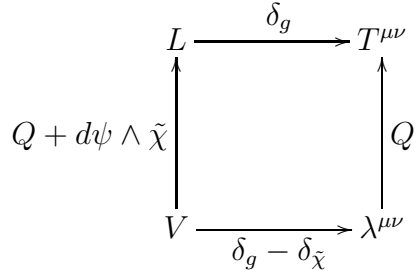
\begin{figure}
    \[
      \renewcommand{\labelstyle}{\textstyle}
      \xymatrix@R=2cm@C=2cm
      { 
      L \ar[r]^{\delta_g} & T^{\mu\nu}\\
        V \ar[r]_{\delta_g-\delta_{\tilde{\chi}}} \ar[u]^{Q + d\psi \wedge \tilde{\chi}} & \lambda^{\mu\nu} \ar[u]_{Q } 
      }
    \]
    \caption{ The $F$-sector relation between $V_2$ and $T_2^{\mu\nu}$ prior to the introduction of the auxiliary field $B_{\mu\nu}$.}
    \label{fig:comdia2}
\end{figure}

Looking at the stress tensor we are led to consider
\begin{equation}
  V_2 = \frac{1}{8} F_{\mu\nu} \tilde{\chi}^{\mu\nu} - \frac{i}{\sqrt{2}} \psi^\mu \partial_\mu \sigma.
\end{equation}
From the form of the supersymmetry variations of $F^+$ and $F^-$ we immediately find a problem here since only $\delta F^-$ is non-zero, but from the self-duality of $\tilde{\chi}$ only the term with $F^+$ survives in the above expression. Thus it seems difficult to generate the term $\tilde{\chi}^{\mu\nu}D_{\mu}\psi_\nu$ in the action. 
This is easily amended by recalling that the six-dimensional theory only gives us information on-shell. This means that it cannot distinguish between the supersymmetry variations for $F^+$ and $F^-$ given in \eqref{eq:twistedsusy} and a (metric independent) variation given by
\begin{equation}
  \delta F_{\mu\nu} = -4 \partial_{[\mu}\psi_{\nu]}v \,,
\end{equation}
which reduces to the ones in \eqref{eq:twistedsusy} using the equation of motion $(\partial_\mu\psi_\nu)^+=0$.
Using the above expression the supersymmetry variation of $V_2$ is given by
\begin{equation}
  \delta_Q V_2 = \frac{i}{2} \partial_\mu \psi_\nu \tilde{\chi}^{\mu\nu} -  \frac{1}{4} F_{\mu\nu} F^{+\mu\nu} - \partial^{\mu}\bar{\sigma} \partial_{\mu}\sigma - i \psi^\mu \partial_\mu \tilde{\eta}.
\end{equation}
The second term can be rewritten as $F_{\mu\nu}F^{+\mu\nu} = \frac{1}{2}F_{\mu\nu}F^{\mu\nu} + \frac{1}{4} \epsilon^{\mu\nu\rho\sigma} F_{\mu\nu}F_{\rho\sigma}$. In this expression the second term is topological and thus will not contribute to the stress-tensor. 
Under a metric variation one then finds that
\begin{equation}
  \delta_g \delta_Q (\sqrt{g} V_2) = \sqrt{g}\,\delta g_{\mu\nu} T^{\mu\nu}_2 + \delta_g\left( \frac{i}{2} \sqrt{g} \partial_\mu \psi_\nu \tilde{\chi}^{\mu\nu}\right).
\end{equation}
The second term on the right-hand side might seem disturbing but it turns out to be crucial for the $Q$-exactness of $T_{\mu\nu}$. This term is zero on-shell but gives a contribution to the metric variation. Rewriting the above expression slightly we have
\begin{equation}
  \sqrt{g}\,\delta g_{\mu\nu} T^{\mu\nu} = \delta_g\delta_Q (\sqrt{g}V_2) -  \delta_g\left(\frac{i}{2}\sqrt{g} \partial_\mu \psi_\nu \tilde{\chi}^{\mu\nu}\right).
  \label{eq:TV2}
\end{equation}
Also here we find that supersymmetry and metric variations do not commute, the first term is not $Q$-exact. However it is the case that
\begin{equation}
  \delta_g \delta_Q (\sqrt{g}V_2) = \delta_Q \Big( \delta_g (\sqrt{g}V_2) - \delta_{\tilde{\chi}} (\sqrt{g}V_2)\Big) + \delta_g \left( \frac{i}{2} \sqrt{g} \partial_{\mu}\psi_\nu \tilde{\chi}^{\mu\nu} \right),
\end{equation}
where $\delta_{\tilde{\chi}}$ denotes the metric variation of the field $\tilde{\chi}$. Notice that the last term is precisely the negative of the last term in \eqref{eq:TV2}.
Using this we find that
\begin{equation}
  \sqrt{g}\delta g_{\mu\nu} T^{\mu\nu} = \delta_Q \Big( \delta_g (\sqrt{g}V_2) - \delta_{\tilde{\chi}} (\sqrt{g}V_2)\Big).
  \label{eq:T2}
\end{equation}
Thus the stress-tensor is indeed $Q$-exact with $T^{\mu\nu}_2 = \{Q, \lambda_2^{\mu\nu}\}$ where 
\begin{equation}
  \sqrt{g}\delta g_{\mu\nu} {\lambda_2^{\mu\nu} = \delta_g (\sqrt{g} V_2) - \delta_{\tilde{\chi}} (\sqrt{g}V_2)}\,,
\end{equation}
in agreement with the relevant terms in the previously derived $\lambda^{\mu\nu}_2$ given in \eqref{eq:lambda2}. Thus, as it stands, the relationship between $V_2$ and $T_2^{\mu\nu}$ can be summarised in Figure \ref{fig:comdia2}.

Following \cite{Baulieu:1988, Brooks:1988, Birmingham:1991} there is an off-shell formulation also for this sector. To ensure the nilpotency of $Q$ we introduce a self-dual auxiliary field $B_{\mu\nu}$ and take
\begin{equation}
  V_2 = \frac{1}{8} \left( F_{\mu\nu} - \frac{1}{2} B_{\mu\nu} \right) \tilde{\chi}^{\mu\nu} - \frac{i}{\sqrt{2}} \psi^\mu D_\mu \sigma\,,
\end{equation}
together with the supersymmetry variations
\begin{equation}
  \begin{aligned}
    \delta \tilde{\chi}_{\mu\nu} &= 2 i v B_{\mu\nu} \\
    \delta B_{\mu\nu} &= 0 \,.
  \end{aligned}
\end{equation}
This achieves two things: firstly the variations now manifestly squares to zero off-shell, but we now also have
\begin{equation}
  \delta_g \delta_Q (\sqrt{g}V_2) = \sqrt{g} \delta g_{\mu\nu} T'^{\mu\nu}_2 \,,
\end{equation}
where $T'^{\mu\nu}_2$ reduces on-shell to $T_2^{\mu\nu}$ in \eqref{eq:T2}. This makes the diagram in Figure \ref{fig:comdia} commute also for this sector.

\section{Conclusions \label{sec:conclusions}}
The topological twisting of abelian (2,0) theory on $C\times M_4$ gives rise, after compactification on $C$, to a four-dimensional Euclidean theory with a  stress tensor  that is $Q$-exact and conserved on a general $M_4$.  
The twisted free theory can be divided into two sectors, one containing the bosonic self-dual two-form consisting of $\{ E_{\mu\nu}, \tilde{\psi}_\mu, A_\mu, \chi_{\mu\nu}, \eta\}$ and one containing the Yang-Mills field strength $\{ F_{\mu\nu}, \tilde{\chi}_{\mu\nu}, \psi_\mu, \sigma, \bar{\sigma}, \tilde{\eta} \}$. The latter  is equivalent to Donaldsson-Witten theory\cite{witten:1988}
which has an off-shell formulation  that can be obtained using the techniques of, e.g.,  \cite{Baulieu:1988, Brooks:1988, Birmingham:1991} 
(see also \cite{spence}).

The  other sector is more subtle however. Here the metric dependence  of the bosonic two-form, and in particular the transformation rules where it enters, make the construction of a $Q$-exact action and a commuting square somewhat complicated. But again it is possible to find a satisfactory set of transformation rules by the introduction of an auxiliary field. The main result is that in the off-shell formulation found here there is a simple relationship between the Lagrangian and  its stress-tensor and the two fermionic quantities, $V$ and $\lambda^{\mu\nu}$, that generate the aforementioned ones under a $Q$ transformation. This relationship can be summarised by a commutative diagram, valid for both off-shell  sectors and displayed
here again for  the benefit  of the reader:
\begin{equation}
      \renewcommand{\labelstyle}{\textstyle}
      \xymatrix@R=2cm@C=2cm{ 
      L  \ar[r]^{\delta_g}  & T^{\mu\nu}\\
         V \ar[r]_{\delta_g} \ar[u]^{Q} &   \lambda_{\mu\nu} \ar[u]_{Q} 
      }
    \label{fig:comdia3}
    \nonumber
\end{equation}

The fermionic quantities $V_i$  are given by
\begin{equation}
  V_1 = -\left( \frac{1}{2} ( i A_\mu + h_\mu) - D_\nu E_\mu{}^\nu \right) \tilde{\psi}^\mu\,,
\end{equation}
for the $E$-sector and
\begin{equation}
  V_2 = \frac{1}{8} \left( F_{\mu\nu} - \frac{1}{2} B_{\mu\nu} \right) \tilde{\chi}^{\mu\nu} - \frac{i}{\sqrt{2}} \psi^\mu D_\mu \sigma\,,
\end{equation}
in the $F$-sector. Here $h_\mu$ and $B_{\mu\nu}$ are the two auxiliary fields needed for the off-shell formulation, the latter being self-dual. The explicit expressions for the remaining 
quantities appearing in the commuting squares of the  two different sectors were presented in the previous section.

The problem previously encountered in \cite{twisted-2-0} regarding the conservation of the stress tensor on a curved background are here alleviated by certain curvature corrections to the bosonic equations of motion. This results in a theory that integrates to an action that is $Q$-exact. 

One can also compare this theory to what one would obtain by first compactifying (2,0) theory on a circle and then compactify once again with a twist down to four dimensions. In the first step we arrive at five-dimensional supersymmetric Yang-Mills. The topological twist of the five-dimensional theory has been investigated \cite{Anderson:2012, Witten:2011} on manifolds of the type $R_+\times M_4$. Even though this is not a compactification of the type we are considering here it is still possible to compare the resulting theories on $M_4$. It is easy to check that a truncated version of the twisted five-dimensional Lagrangian corresponds to the one presented here.
In light of these results it would be interesting to see how the theory depends on the higher modes on $C$. There are arguments\cite{Yagi:1, Witten:2011} for why the theory on $C$ should become holomorphic after the full twist in the Euclidean setup and it would therefore be of interest to see if and how this manifests itself here.

\section*{Acknowledgement}
HL would like to thank Jakob Palmkvist and Louise Anderson for useful discussions.
UG and HL are supported by the Knut and Alice Wallenberg Foundation and BN is partly funded by the Swedish Research Council.

\appendix
\section{Metric variation of self-dual forms}
When performing metric variations in a theory with self-dual two-forms it is important that their  self-duality is preserved. To see how this is done we consider the self-duality constraint
in the form \begin{equation}
  A^+ = \star A^+,
\end{equation}
where $A^+$ is a two-form and $\star$ is the Hodge dual. From the metric dependence of the Hodge dual we see that $A^+$ fails to be self-dual under the perturbed metric $g+\delta g$. Let us then assume that $A^+\rightarrow A^++\delta_g A^+$ as $g\rightarrow g+\delta g$. In the perturbed metric the self-duality condition  then reads
\begin{equation}
  \delta_g A^+ = (\delta_g \star) A^+ + \star \delta_g A^+,
\end{equation}
which is equivalent to
\begin{equation}
  \frac{1}{2}(1-\star)\delta_g A^+ = \frac{1}{2}(\delta_g \star) A^+.
\end{equation}
To preserve the relation $A^+ = \star A^+$ under metric variations we must therefore impose the anti-self-dual variation above\footnote{One easily verifies that the right-hand side is anti-self-dual using the self-duality of $A^+$ and the fact that $\star^2 = 1$ implies $\delta_g \star \star = -\star \delta_g \star$.}. Since the self-dual part of $\delta_g A^+$ is unconstrained we can take it to vanish.
In components the variation above takes the form
\begin{equation}
  \delta_g A^+_{\mu\nu} = \frac{1}{2} \delta g^{\rho \rho'} \epsilon_{\mu \nu \rho}{}^{\sigma} A^+_{\rho' \sigma} - \frac{1}{8} \delta g^{\lambda \tau} g_{\lambda \tau} \epsilon_{\mu \nu}{}^{\rho \sigma} A^+_{\rho \sigma}.
\end{equation}

\bibliography{refs}{}

\providecommand{\href}[2]{#2}\begingroup\raggedright\begin{thebibliography}{10}

\bibitem{twisted-2-0}
L.~Anderson and H.~Linander, {\it {The trouble with twisting (2,0) theory}},
  {\em JHEP} {\bf 1403} (2014) 062,
  [\href{http://xxx.lanl.gov/abs/1311.3300}{{\tt arXiv:1311.3300}}].

\bibitem{Yagi:1}
J.~Yagi, {\it {On the Six-Dimensional Origin of the AGT Correspondence}},  {\em
  JHEP} {\bf 1202} (2012) 020, [\href{http://xxx.lanl.gov/abs/1112.0260}{{\tt
  arXiv:1112.0260}}].

\bibitem{AGT}
L.~F. Alday, D.~Gaiotto, and Y.~Tachikawa, {\it {Liouville Correlation
  Functions from Four-dimensional Gauge Theories}},  {\em Lett.Math.Phys.} {\bf
  91} (2010) 167--197, [\href{http://xxx.lanl.gov/abs/0906.3219}{{\tt
  arXiv:0906.3219}}].

\bibitem{Gaiotto:N2dualities}
D.~Gaiotto, {\it {N=2 dualities}},  {\em JHEP} {\bf 1208} (2012) 034,
  [\href{http://xxx.lanl.gov/abs/0904.2715}{{\tt arXiv:0904.2715}}].

\bibitem{Witten:2011}
E.~Witten, {\it {Fivebranes and Knots}},
  \href{http://xxx.lanl.gov/abs/1101.3216}{{\tt arXiv:1101.3216}}.

\bibitem{Dimofte:2011ju}
T.~Dimofte, D.~Gaiotto, and S.~Gukov, {\it {Gauge Theories Labelled by
  Three-Manifolds}},  {\em Commun.Math.Phys.} {\bf 325} (2014) 367--419,
  [\href{http://xxx.lanl.gov/abs/1108.4389}{{\tt arXiv:1108.4389}}].

\bibitem{Kapustin:2006pk}
A.~Kapustin and E.~Witten, {\it {Electric-Magnetic Duality And The Geometric
  Langlands Program}},  {\em Commun.Num.Theor.Phys.} {\bf 1} (2007) 1--236,
  [\href{http://xxx.lanl.gov/abs/hep-th/0604151}{{\tt hep-th/0604151}}].

\bibitem{vafa-witten}
C.~Vafa and E.~Witten, {\it {A Strong coupling test of S duality}},  {\em
  Nucl.Phys.} {\bf B431} (1994) 3--77,
  [\href{http://xxx.lanl.gov/abs/hep-th/9408074}{{\tt hep-th/9408074}}].

\bibitem{Witten:1995}
E.~Witten, {\it {Some comments on string dynamics}},
  \href{http://xxx.lanl.gov/abs/hep-th/9507121}{{\tt hep-th/9507121}}.

\bibitem{Witten:ct46}
E.~Witten, {\it {Conformal Field Theory In Four And Six Dimensions}},
  \href{http://xxx.lanl.gov/abs/0712.0157}{{\tt arXiv:0712.0157}}.

\bibitem{witten:1988}
E.~Witten, {\it {Topological Quantum Field Theory}},  {\em Commun.Math.Phys.}
  {\bf 117} (1988) 353.

\bibitem{yamron:1988}
J.~P. Yamron, {\it {Topological Actions From Twisted Supersymmetric Theories}},
   {\em Phys.Lett.} {\bf B213} (1988) 325.

\bibitem{marcus:1995}
N.~Marcus, {\it {The Other topological twisting of N=4 Yang-Mills}},  {\em
  Nucl.Phys.} {\bf B452} (1995) 331--345,
  [\href{http://xxx.lanl.gov/abs/hep-th/9506002}{{\tt hep-th/9506002}}].

\bibitem{Baulieu:1988}
L.~Baulieu and I.~Singer, {\it {Topological Yang-Mills Symmetry}},  {\em
  Nucl.Phys.Proc.Suppl.} {\bf 5B} (1988) 12--19.

\bibitem{Brooks:1988}
R.~Brooks, D.~Montano, and J.~Sonnenschein, {\it {Gauge Fixing and
  Renormalization in Topological Quantum Field Theory}},  {\em Phys.Lett.} {\bf
  B214} (1988) 91.

\bibitem{Birmingham:1991}
D.~Birmingham, M.~Blau, M.~Rakowski, and G.~Thompson, {\it {Topological field
  theory}},  {\em Phys.Rept.} {\bf 209} (1991) 129--340.

\bibitem{lee-lee-park}
K.~Lee, S.~Lee, and J.-H. Park, {\it {Topological Twisting of Multiple M2-brane
  Theory}},  {\em JHEP} {\bf 0811} (2008) 014,
  [\href{http://xxx.lanl.gov/abs/0809.2924}{{\tt arXiv:0809.2924}}].

\bibitem{Anderson:2012}
L.~Anderson, {\it {Five-dimensional topologically twisted maximally
  supersymmetric Yang-Mills theory}},  {\em JHEP} {\bf 1302} (2013) 131,
  [\href{http://xxx.lanl.gov/abs/1212.5019}{{\tt arXiv:1212.5019}}].

\bibitem{Seiberg:noteon16}
N.~Seiberg, {\it {Notes on theories with 16 supercharges}},  {\em
  Nucl.Phys.Proc.Suppl.} {\bf 67} (1998) 158--171,
  [\href{http://xxx.lanl.gov/abs/hep-th/9705117}{{\tt hep-th/9705117}}].

\bibitem{spence}
B.~J. Spence, {\it {Topological Born-Infeld actions and D-branes}},
  \href{http://xxx.lanl.gov/abs/hep-th/9907053}{{\tt hep-th/9907053}}.

\end{thebibliography}\endgroup
\bibliographystyle{JHEP}

\end{document}